**Pressure-Temperature Phase Diagram of Vanadium Dioxide**

Yabin Chen,[1] Shuai Zhang,[2] Feng Ke,[3] Changhyun Ko,[1] Sangwook Lee,[1] Kai Liu,[4] Bin Chen,[3] Joel W. Ager,[1,5] Raymond Jeanloz,[2] Volker Eyert,[6] and Junqiao Wu[1,5,*]

1. Department of Materials Science and Engineering, University of California, Berkeley, CA 94720, USA.
2. Department of Earth and Planetary Science, University of California, Berkeley, CA 94720, USA.
3. Center for High Pressure Science and Technology Advanced Research, Shanghai 201203, China.
4. State Key Laboratory of New Ceramics and Fine Processing, School of Materials Science and Engineering, Tsinghua University, Beijing 100084, China.
5. Materials Sciences Division, Lawrence Berkeley National Laboratory, Berkeley, CA 94720, USA.
6. Materials Design SARL, 92120 Montrouge, France.

*Corresponding author: wuj@berkeley.edu
Y.C., S.Z., and F.K. contributed equally to this work.

**Abstract**

The complexity of strongly correlated electron physics in vanadium dioxide is exemplified as its rich phase diagrams of all kinds, which in turn shed light on the mechanisms behind its various phase transitions. In this work, we map out the hydrostatic pressure - temperature phase diagram of vanadium dioxide nanobeams by independently varying pressure and temperature with a diamond anvil cell. In addition to the well-known insulating M1 (monoclinic) and metallic R (tetragonal) phases, the diagram identifies the existence at high pressures of the insulating M1' (monoclinic, more conductive than M1) phase, and two metallic phases of X (monoclinic) and O (orthorhombic, at high temperature only). Systematic optical and electrical measurements combined with density functional calculations allow us to delineate their phase boundaries as well as reveal some basic features of the transitions.

Vanadium dioxide (VO$_2$), as a prototype of strongly correlated electron materials, is of great interest in condensed matter physics as well for device applications.[1-3] The electron correlation interplaying with lattice stabilizes a very rich phase diagram of VO$_2$ that consists of many phases with distinct structures and electronic properties. Transitions between these phases can be driven by temperature, photo-excitation, hydrostatic pressure, uniaxial stress, or electrical gating.[1, 4-12] For example, the well-known, insulating M$_1$ (monoclinic, space group *P2$_1$/c*) phase transforms to the metallic R (tetragonal, *P4$_2$/mnm*) phase at 68 °C under ambient pressure.[1] This M1-R transition can also be driven by hydrostatic pressure[6] or uniaxial compression.[4] Application of hydrostatic pressure is a clean and powerful way to tune the lattice and electronic degrees of freedom. Hydrostatic pressure applied via a diamond anvil cell (DAC) has also been able to drive the M1 phase to another isostructural, more conductive M1' phase, and finally a new metallic X (monoclinic) phase at room temperature, and to drive the metallic R phase to new metallic O and X phases at 383 K.[9] A pressure-temperature phase diagram has been proposed for VO$_2$.[8] However, this phase diagram with three regions of M1, Mx and R is largely schematic, and their phase boundaries, especially at high pressures, are unclear, undefined and unsubstantiated with experimental data. In this work, we establish a clear hydrostatic pressure-temperature ($P$ - $T$) phase diagram for VO$_2$, quantitatively delineating the phase boundaries between the M1, M1', R, O, and X phases, and provide comparisons to first-principles calculations.

The phase transitions were identified within a heated DAC by various techniques, including Raman spectroscopy, optical reflectance, and electrical transport. The VO$_2$ nanobeams used in this work were grown using a vapor transport method.[13] They typically have a flat surface and small (<1 μm) thickness, and exhibit a single-domain, sharp metal-insulator transition (MIT) at $T_{MIT}$ = 68 °C, with a narrow hysteresis. All high-pressure experiments were performed in a four-post DAC with 300 μm culet size (see Supporting Information). The pressure-transmitting medium was a mixture of methanol-ethanol (4:1) and Daphne 7373, for Raman and electrical measurements, respectively. A quasi four-probe geometry was used for the electrical measurements using Pt foil as electrodes, and the micro-device fabrication method can be found elsewhere.[14] High temperatures were realized by heating the DAC in a water bath or with heating belts. The pressure was calibrated with the standard method of ruby photoluminescence.[15] The temperature was measured with a thermocouple placed very close to the gasket chamber. Theoretical calculations were performed within the framework of density functional theory (DFT). Electron-electron exchange and correlation effects were described by the generalized gradient approximation with the Perdew-Burke-Ernzerhof (PBE)[16] functional plus Hubbard $U$.[17] The density-functional or Kohn-Sham equations were solved with the plane-wave pseudopotential method, as implemented in Vienna *ab initio* Simulation Package.[18, 19] The $k$-mesh for sampling the Brillouin zone and plane-wave cutoff energy for expanding the electronic wavefunctions were chosen such that the total energies were converged within 0.65 meV/formula unit (f.u.). We chose $U$ = 3.4 eV, which produces the bandgap of M1 phase (0.7 eV) correctly.[20]

Figure 1 shows the phase transition driven by temperature under constant pressure (*i.e.*, isobaric condition). The pressure was fixed at 9.6 GPa for experiments in Figure 1. As shown in Figure 1a, a sharp change in optical reflection is seen within a narrow temperature range (< ~0.2 °C), signifying the transition from the insulating (bright) phase to the metallic (dark) phase, consistent with previous observations on this type of specimens.[21,22] The very narrow temperature range further confirms the nearly single-domain nature of the phase transition. This is also an indication that the specimens, which are transferred from the as-grown substrate to the diamond surface, sit loosely on the surface instead of being firmly clamped on it, and as such, any strain imposed by the diamond surface onto the specimen is negligible. The lack of strong clamping is also beneficial for the molecules of the pressure medium to sip in underneath the sample, thus ensuring the hydrostaticity of the pressure transmitted to the sample. Raman spectra of the sample obtained at these temperatures are shown in Figure 1b. From the Raman spectrum, the low-temperature phase is clearly identified as the M1 phase. Combining isotope substitution and density functional theory calculations, it has been established that in the M1 phase, the two low-frequency peaks of $\omega$ = 193.9 and 228.5 cm$^{-1}$ (at 21.6 °C) correspond to V-V lattice vibration, and all other Raman peaks are related to V-O vibration modes.[23] All these modes slightly soften with increasing temperature, until reaching the transition temperature, above which all Raman peaks disappear owing to the metallicity of the R phase. The transition temperature obtained by Raman is identical with that obtained by optical reflection. More importantly, the transition temperature of $T_{MIT}$ = 77.4 °C at 9.6 GPa is higher than $T_{MIT}$ = 68 °C at ambient pressure (~0 GPa). This is consistent with previous report of a positive slope in the M1-R phase boundary.[6]

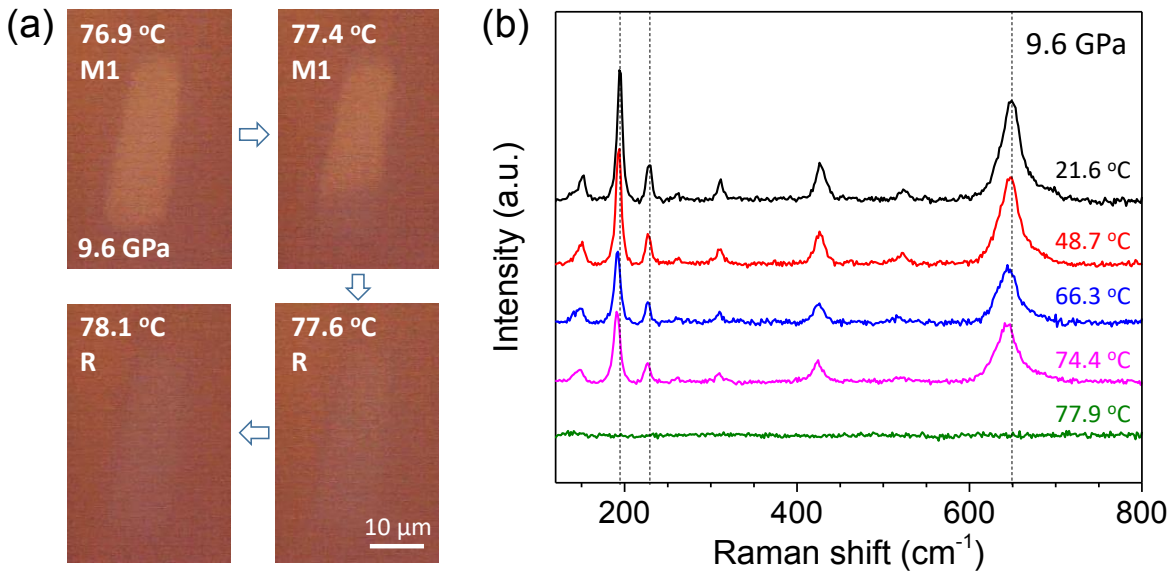

**Figure 1.** Phase transition of $VO_2$ driven by temperature under isobaric condition. (a) Optical images of a $VO_2$ nanobeam across the M1-R phase transition at fixed pressure of 9.6 GPa. M1 and R phases show bright-yellow and dark-green contrast, respectively. (b) Raman spectrum evolving with temperature at the fixed pressure of 9.6 GPa.

At constant temperature (*i.e.*, isothermal condition), hydrostatic pressure also drives $VO_2$ from the M1 phase to new phases. Shown in Figure 2a is the Raman spectra taken at fixed temperature of 50 °C with increasing pressure. The Raman peaks of the M1 structure gradually shift to higher wavenumbers with pressure, until very high pressure (~33 GPa) where they all disappear, indicating a metallic phase at and beyond that pressure. We mark this metallic phase as X phase following previous work.[9] Plotting the pressure dependence of the three most prominent Raman peaks reveals new information, as shown in Figure 2b. It can be seen that the V-O mode at $\omega = 617.5$ cm$^{-1}$ (2.5 GPa) blue shifts linearly with an initial rate of $d\omega/dP = 4.1$ cm$^{-1}$/GPa, but then the rate changes to a much shallower, 2.4 cm$^{-1}$/GPa when the pressure is higher than 16.5 GPa. The two V-V modes, on the other hand, show no obvious change in the slope ($d\omega/dP$). However, we note that at lower temperatures such as room temperature (Supporting information, Figure S2), these two V-V modes also show slight slope change at the same pressure where the V-O mode shows the slope change. This is consistent with previous, room-temperature pressure studies.[7,23] Therefore, before reaching the metallic phase, all the M1 Raman modes remain, but some (or all, depending on temperature) of them show a change in the pressure rate at ~16.5 GPa. This effect points to a continuous phase transition from M1 to a new, insulating M1' phase at this pressure, as suggested also by our previous room-temperature micro-x-ray diffraction (XRD) experiments.[9] The M1' phase is isostructural to M1, yet with very different electrical resistance as discussed below.

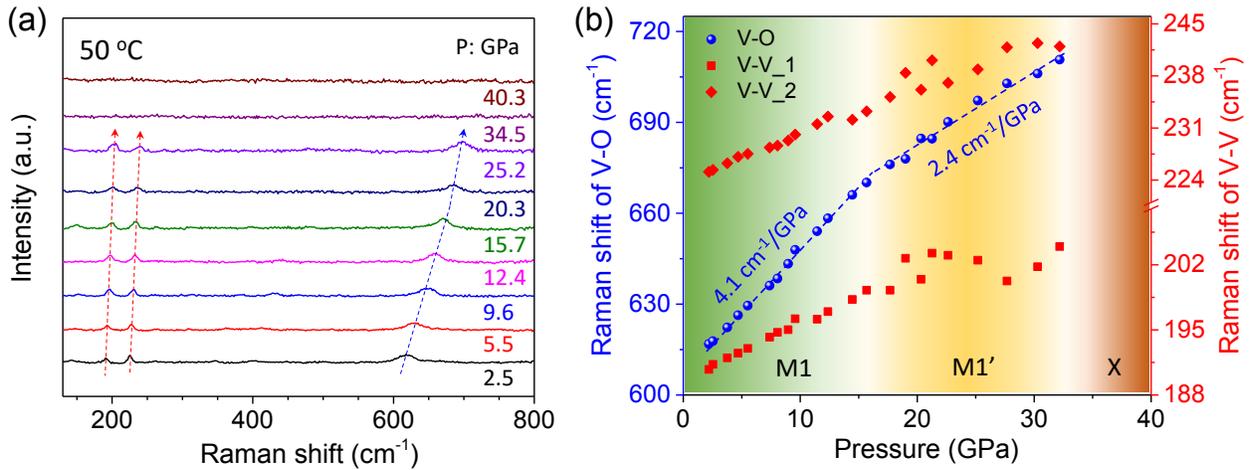

**Figure 2.** Phase transition of VO$_2$ driven by pressure under isothermal condition. (a) Raman spectrum evolving from M1 to M1' and then X phase at fixed temperature of 50 °C. (b) Peak positions of different Raman modes as a function of pressure extracted from (a).

In order to probe the electronic nature of these phases, electrical transport measurements were also performed as a function of pressure or temperature. Figure 3a shows the resistance (R) - temperature (T) dependence at fixed pressures. Similar to typical transport measurements at ambient pressure, at low temperatures VO$_2$ behaves as a doped semiconductor, with a decreasing resistance with temperature. At certain temperature ($T_{MIT}$), the resistance abruptly drops down by several orders of magnitude. The process is reversible, as cooling at the same pressure shows a very similar R-T dependence with a small hysteresis as seen in Figure 3a. There are some small features on the R-T curves, possibly attributed to inhomogeneous pressure in the DAC. The resistance was also measured as a function of pressure at room temperature, as shown in Figure 3b. A drastic change in the pressure dependence is seen in the range of 15~20 GPa, which is about the same pressure where the Raman peaks change slope in Figure 2b. Therefore, the much more rapidly decreasing resistance in Figure 3b is associated with the M1' phase. Such a new phase at high pressures has been considered as the onset of metallization process occurring in the monoclinic structure.[7,24] Our observation of the M1-like Raman peaks together with rapidly decreasing, yet non-metallic, resistance in the M1' phase shows that it is still a monoclinic-structured insulating phase, but with a narrow bandgap rapidly closing from that of M1 phase ($E_g \approx 0.7$ eV).[20] As pressure further increases beyond ~ 38 GPa, VO$_2$ transforms into the metallic X phase showing very low resistance.

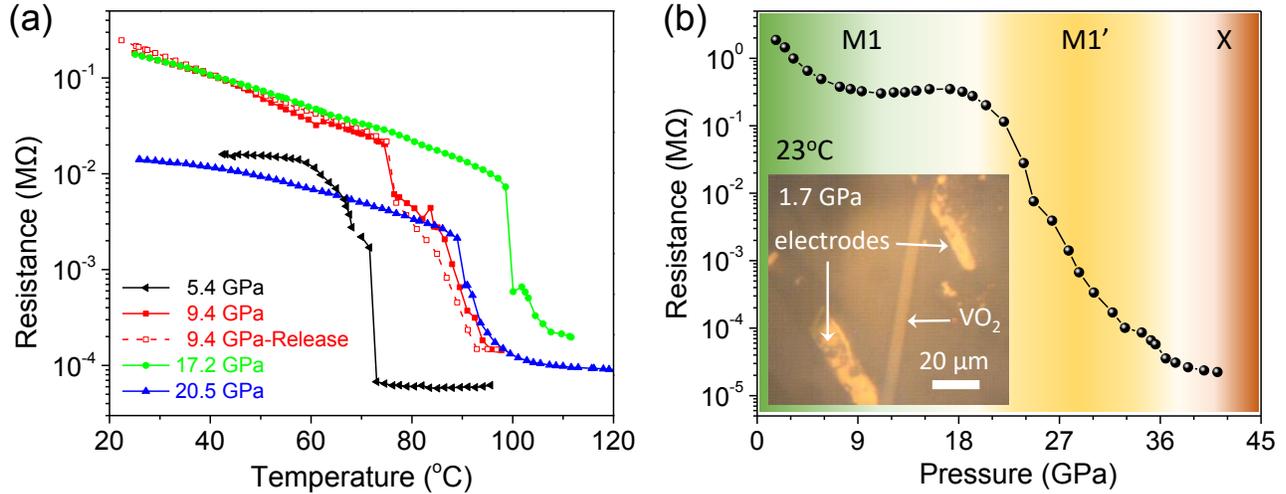

**Figure 3.** Electrical resistance of VO$_2$ across the phase transitions driven by temperature or pressure. (a) Temperature dependence of resistance of VO$_2$ under isobaric condition. (b) Pressure dependence of resistance of VO$_2$ at room temperature. Inset is optical image of a quasi-four probe device for electrical measurements.

The optical reflectance, Raman, and electrical transport data obtained in isothermal and/or isobaric conditions allow us to construct a pressure - temperature phase diagram of VO$_2$, including M1 (I), M1' (I), R (M), O (M), and X (M) phases as shown in Figure 4. Four phase transition points (black squares) obtained with micro-XRD at room temperature as well as 110 °C are also added for comparison.[9] It can be seen from Figure 4 that, although these data points were taken with different techniques (Raman, electrical, reflectance or XRD), along different axes of variables (isobaric or isothermal), and on different samples, they clearly define a consistent *P-T* phase diagram over a wide pressure and temperature range. This phase diagram share some similarities to the largely schematic diagram of phases reported in Ref.[8]: the *P-T* phase boundary has a positive slope at low pressures, and the evidence of new metallic phases at high pressures. However, the most striking features of this diagram are: 1) the dome-like, non-monotonic phase boundary between the low-temperature, I phases and high-temperature, M phases (*i.e.*, the MIT boundary), and 2) the gradual evolution along the pressure axis from the insulating M1 to the less insulating M1' and finally the metallic X phases, while the M1 - M1' boundary incidentally extrapolates to the vertex of the MIT dome. In the following we analyze these features.

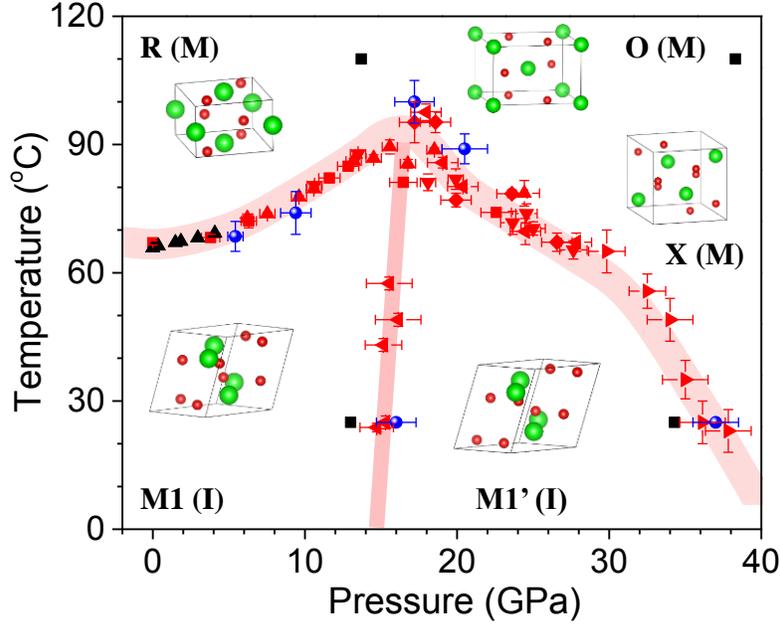

**Figure 4.** Phase diagram of $VO_2$ over a board *P-T* range, including M1 (I), M1' (I), R (M), O (M), and X (M). I and M represent insulating and metallic phases, respectively. Insets are the simulated atomic structures of each phases of $VO_2$ (V: green atoms, O: red atoms). The red and blue data points on the phase boundaries are obtained from Raman and electrical measurements, respectively, and different shapes of symbols represent different samples. The black square data points are determined by XRD extracted from Ref.[9] The black triangle data points are from Ref.[6] The error bars include the temperature and pressure uncertainties. The broad lines defining the phase boundaries are guide to the eye.

    For the transition from the well-known M1 (I) to R (M), the slope of the phase boundary d*T*/d*P* is determined to be 1.4 °C/GPa when averaged over a wide range of *P* from 0 to 16 GPa. This is higher than the reported d*T*/d*P* = 0.8 °C/GPa measured over *P* < 4 GPa.[6] It can be seen from Figure 4 that the M1-R phase boundary takes a convex line shape, showing higher slope at higher pressures. The M1-R transition is a first-order phase transition, with a discontinuous change in volume of $\Delta V/V$ = 0.044% at ambient pressure.[25] These quantities are related through the Clausius-Clapeyron equation d*T*/d*P* = $T\Delta V/L$, where $\Delta V = V_R - V_{M1}$ is the molar volume change and *L* is the molar enthalpy change (latent heat) going from M1 to R. The positive slope (d*T*/d*P*) is an indication of volume expansion going from M1 to R, consistent with reports.[8, 25] Using the average d*T*/d*P* = 1.4 °C/GPa and the ambient $\Delta V/V$ = 0.044%, an average *L* is calculated to be ~1.9 kJ/mol, compared to averaged value of *L* = 3.1~4.3 kJ/mol in literature.[25]

    By considering the *P* dependencies of d*T*/d*P* and $\Delta V$, it is possible to go beyond these averaged values and obtain values of *L* under pressure. The molar volume, *V*, has been experimentally measured as a function of *P* at 25 °C (M1 phase) and 110 °C (R phase) in Ref.,[9] as well as a function of *T* (both M1 and R phases) at ambient pressure in Ref.,[25] and these dependencies all appear linear within the range of pressure and temperature variations. Assuming that both $V_{M1}$ and $V_R$ depend on *P* and *T* following the linear functional form of $\alpha \cdot P + \beta \cdot T + \delta$, by fitting to the experimental data in Refs.,[9,25] one can obtain $\alpha_{M1}$, $\beta_{M1}$, $\delta_{M1}$ and $\alpha_R$, $\beta_R$, $\delta_R$.

Consequently, the volume change $\Delta V = V_R - V_{M1}$ is obtained as a function of $T$ and $P$, and its values along the M1-R phase boundary is determined. Combining the values of $\Delta V$ with the slope $dT/dP$ of the measured phase boundary shown in Figure 4, variation of $L$ is obtained along the phase boundary. As shown in Figure S3c, $L$ increases as a function of pressure, varying from 1.3 kJ/mol at ambient pressure to 12.2 kJ/mol at ~ 16 GPa. Budai *et al.*[26] showed that the R structure is much more anharmonic than the M1 phase, and that the total entropy change (0.43 $k_B$/atom) across the MIT at ambient pressure is mostly attributed to softer anharmonic phonons in the R phase, rather than electronic contribution. Our discovery of increasing $L$ points to even stronger lattice anharmonicity in the R structure at higher pressures.

We performed DFT calculations to investigate the M1'(I) - X(M) transition. The calculations for these phases covered pressures from 10 to 90 GPa as shown in Figure S4. By full optimization of the structure, the enthalpy and lattice parameters of the two phases were obtained. The volume is calculated to decrease by ~5% from that of the M1' to X phase, which is much higher than that of the M1-R transition. As the slope $dT/dP$ of the M1'-X phase boundary is negative with an averaged value on the same order of magnitude with that of the M1-R boundary, the average latent heat $L$ is estimated to be positive going from M1' to X, and much higher than $L$(M1-R), reaching ~155 kJ/mol, suggesting much higher entropy in the metallic X phase than in the metallic R phase.

For the M1 to M1' transition, the phase boundary is nearly vertical. We calculated the pressure-driven change in phonon frequencies using the finite differences method (see Figure S5). At $P > 15$ GPa, some phonon modes start to deviate substantially in pressure dependency from those of the M1 phase. We evaluated the room-temperature Grüneisen parameters of the Raman modes, $\gamma = -\partial \ln(\omega)/\partial \ln(V) = -(V/\omega)(d\omega/dP)/(dV/dP)$. As shown in Figure S2, for the V-V Raman mode at 222.1 cm$^{-1}$ (0 GPa), $d\omega/dP$ increases from 0.2 for M1 to 0.8 cm$^{-1}$/GPa for the M1' phase, while for the V-O mode at 608.2 cm$^{-1}$ (0 GPa), it decreases from 4.2 for M1 to 1.7 cm$^{-1}$/GPa for M1'. Using previous results of experimentally measured $dV/dP$,[9] the obtained $\gamma$ values exhibit an abrupt change at the M1-M1' transition pressure, with a sharp increase in $\gamma$(V-V) and decrease in $\gamma$(V-O).

Finally, for the transition from the metallic R to the metallic O phase, calculations predict an onset of the transition in the range of 12 ~ 14 GPa (see Figure S6). As expected, the O phase comes with a gradually reduced volume (by 0.3% at 14 GPa and 2.2% at 30 GPa) as compared to the R-phase unit cell. Meanwhile, the *a*- and *b*-lattice parameters split into two values above the pressure for the tetragonal structure. Note that the absolute shifts of the atoms are even more pronounced, since the oxygen parameters are given relative to the cell parameters and thus have to be multiplied by these in order to obtain the real distance between atoms. This R - O phase transition is also signaled by the difference of the enthalpies of the R and O phases as displayed in Figure S6a. The enthalpy calculated for the O structure starts to lower at about 13 GPa, and then experiences a drastic drop between 15 and 30 GPa, thus stabilizing the O structure. We tried other functionals (PBE, and PBEsol[27]), and found the same energy and structural relations between R and O phases.

In conclusion, we have experimentally mapped the phase diagram of VO$_2$ by independently varying pressure and temperature in a diamond anvil cell, which is also analyzed and supported by first-principles calculations. The diagram establishes phase boundaries not only between the well-known, insulating M1 phase and the metallic R phase, but also between the M1 phase and a more conductive M1' phase at intermediate pressures, as well as between the M1' phase and a metallic X phase at very high pressures. At high temperatures, the existence of another metallic O phase is also discussed. Our established phase diagram may serve as an instructive benchmark for ultimate elucidation of the phase transition physics of $VO_2$, as well as its potential applications.

**Acknowledgements**

This work was supported by U.S. NSF Grant No. 1608899. Y.C., J.W., and J.W.A. acknowledge support from the Singapore-Berkeley Research Initiative for Sustainable Energy (SinBeRISE). The laser milling was supported by BL12.2.2 at the Lawrence Berkeley National Laboratory and COMPRES (Grant No. EAR 11- 57758). The authors acknowledge useful discussions with Dr. Hong Ding on DFT simulation, and Dr. Wen Fan on DAC experiments.

**Supporting Information.**

Details of $VO_2$ growth and device fabrication in the DAC; Raman of a $VO_2$ nanobeam under isobaric heating; Raman during M1 to M1' transition driven by isothermal compression at room temperature; Volume change and latent heat of M1 to R transition; Calculated M1' to X transition; Calculated phonon frequencies as a function of pressure for the M1 - M1' transition; Calculated R to O transition; Comparison of calculated enthalpy between M1, M1', R, O, and X phases at different pressures.